\documentclass[preprint,showpacs,prb,floatfix,superscriptaddress,
citeautoscript ,cite]{revtex4}
\usepackage{graphicx}
\usepackage{color}
\usepackage{amsmath}

\usepackage{booktabs}

\begin{document}

\title{Mg(OH)$_2$-WS$_2$ Heterobilayer: Electric Field Tunable Bandgap
Crossover}

\author{M. Yagmurcukardes}
\email{mehmetyagmurcukardes@iyte.edu.tr}
\affiliation{Department of Physics, Izmir Institute of Technology, 35430 Izmir,
Turkey}

\author{E. Torun}
\affiliation{Department of Physics, University of Antwerp, Groenenborgerlaan
171, B-2020 Antwerp, Belgium}

\author{R. T. Senger}
\email{tugrulsenger@iyte.edu.tr }
\affiliation{Department of Physics, Izmir Institute of Technology, 35430 Izmir,
Turkey}

\author{F. M. Peeters}
\affiliation{Department of Physics, University of Antwerp, Groenenborgerlaan
171, B-2020 Antwerp, Belgium}

\author{H. Sahin}
\affiliation{Department of Physics, University of Antwerp, Groenenborgerlaan
171, B-2020 Antwerp, Belgium}

\date{\today}

\begin{abstract}
Magnesium hydroxide (Mg(OH)$_2$) has a layered brucite-like structure in its
bulk form and was recently isolated as a new member of 2D monolayer
materials. We investigated the electronic and optical
properties of monolayer crystals of Mg(OH)$_2$ and WS$_2$ and their possible
heterobilayer structure by means of first principles
calculations. It was found that both monolayers of Mg(OH)$_2$ and WS$_2$
are direct-gap semiconductors and these two monolayers
form a typical
van der Waals heterostructure with a weak interlayer interaction and
a type-II band alignment with a staggered
gap
that spatially seperates electrons and holes. We also showed that an
out-of-plane electric field
induces
a transition from a staggered to a straddling type heterojunction.
Moreover, by solving the Bethe-Salpeter equation on top of single
shot
G$_0$W$_0$ calculations, we show that the oscillator strength of the
intralayer excitons of the heterostructure is an order of
magnitude larger than the oscillator strength of the interlayer excitons.
Because of the staggered interfacial gap and the field-tunable energy band
structure,
the Mg(OH)$_2$-WS$_2$ heterobilayer can become an
important
candidate for various optoelectronic device applications in nanoscale.

\end{abstract}

\maketitle

\section{Introduction}

Over the past decade, graphene, a two dimensional form of carbon atoms arranged
in a honeycomb
structure, led to an enormous interest in the field of two
dimensional materials due to its exceptional physical
properties\cite{Novo1,Geim1}.
However, the lack of a band gap is a major obstacle for the use of graphene in
optoelectronic applications. Subsequently other novel two dimensional (2D)
materials such as hexagonal structures of III-V binary
compounds\cite{Novo3,hasan1} and transition metal dichalcogenides
(TMDs)\cite{Gordon,Wang} have gained a lot of interest due to
their wide range of band gap energies. The synthesized members of TMDs, notably
MoS$_2$\cite{Radisavlijevic}, MoSe$_2$\cite{Lu},
WS$_2$\cite{Matte}, and recently ReS$_2$\cite{Tongay} and
ReSe$_2$\cite{Wolverson} which
have band gaps around 1-2 eV, are suitable monolayer materials for many
optoelectronic  applications. Beyond being novel atomic-thick materials,
lateral and vertical heterostructures of these monolayer crystals have also
recieved considerable
attention.

As constituents of possible heterostructures TMDs are very promissing. Those new
members of 2D monolayer materials have tunable
electronic properties from metalic to wide-gap semiconducting\cite{Wilson,Ataca}
and excellent mechanical
properties\cite{Gomez}. Moreover, TMDs can be used in various
fields such as nanoelectronics\cite{Radisavlijevic,Li,Popov},
photonics\cite{Eda,Mak,Yin}, and for transistors\cite{Wang},
catalysis\cite{Drescher},
hydrogen storage\cite{Seayad}, and Li-ion battery applications\cite{Chang}.
Among TMDs, WS$_2$ has been studied intensively. It is an indirect-gap
semiconductor in its bulk form while it shows a transition to direct-gap
character in its monolayer form\cite{Boker,Klein,Thomalla}. It was shown
by
Ramasubramaniam that the optoelectronic
properties of WS$_2$ and MoX$_2$ (X=S or Se) monolayers are tunable
through quantum confinement of
carriers within the monolayers\cite{Ramasubramaniam}. Shi \textit{et al.}
showed that the electron effective mass
decreases
as the applied strain increases, and monolayer
WS$_2$ possesses the lightest charge carriers among the TMDs\cite{Shi}. In
addition, strong excitonic features of WS$_2$, including neutral and redshifted
charged excitons were observed by Mak \textit{et al.}\cite{Mak2}
Due to these interesting electronic and optical properties, one may go a step
further and construct 2D
heterostructures incorporating monolayer WS$_2$ with other 2D monolayer with
the potential to achieve
enhanced functionalities.

Recently synthesized monolayer of Mg(OH)$_2$, a member of
alkaline-earth hydroxides (AEH), with formula X(OH)$_2$ where X = Mg or Ca, are
candidate materials for constructing such heterostructures. Magnesium and
calcium
hydroxides are
multifunctional materials which
have many important applications in industry, technology, solid-state
electronics, and in photovoltaic devices\cite{Estrela,Ghali,Cao,Snider}.
Recently, we studied Ca(OH)$_2$ monolayer crystals and found that the
number of layers of Ca(OH)$_2$ does not affect the electronic, structural, and
magnetic properties
qualitatively while the intrinsic mechanical stiffness of
each layer becomes slightly larger as the structure changes
from monolayer to bilayer.
Very recently, Torun \textit{et al.}\cite{Torun} investigated
the electronic and optical properties of the heterobilayer structure
GaS-Ca(OH)$_2$ and found that it is a type-II
heterojunction where spatially
seperated charge carriers can be formed. The optical spectra of different
stacking types exhibit distinct properties. Like Ca(OH)$_2$,
Mg(OH)$_2$ has a layered
structure in its bulk form possessing the trigonal symmetry of the space group
P$\overline{3}$m1 (brucite)\cite{Desgranges,Catti,Busing}. Mg(OH)$_2$ itself is
a wide-gap insulator with a band gap
of 7.6 eV found experimentally for the bulk structure\cite{Murakami}. Kuji
\textit{et al.} reported properties of C-doped Mg(OH)$_2$ films and found that
the material becomes transparent in the visible region and electrically
conducting which are favourable properties for applications in photovoltaic
devices\cite{Kuji}. Huang \textit{et
al.}\cite{Huang} found experimentally a spectral peak near the band edge
corresponding to
strongly localized excitons with an exciton binding energy of 0.53 eV. This
indicates a strong localization of
the hole and electron to the oxygen p$_x$ and p$_y$ states. Most recently,
successful synthesis of Mg(OH)$_2$ monolayers on MoS$_2$ and their optical
properties were reported by Suslu \textit{et
al.}\cite{Suslu}

Here, we predict an electric field dependence of the electronic and optical
properties of the Mg(OH)$_2$-WS$_2$ heterobilayer structure. Our results reveal
that monolayer crystal of Mg(OH)$_2$ combined with TMDs may lead to the
emergence
of novel multifunctional nanoscale optoelectronic devices.

The paper is organized as
follows: Details of the computational methodology is given in Sec. \ref{comp}.
Structural and electronic properties of monolayers of Mg(OH)$_2$ and WS$_2$ are
presented in Sec. \ref{monolayer} while the structural properties of the
Mg(OH)$_2$-WS$_2$ heterobilayer are presented in Sec. \ref{hetero}. The
effect of an external electric field on the electronic properties of the
heterobilayer structure is given in Sec. \ref{Efield}. In Sec. \ref{optical}
the electric field dependence of the optical properties of the heterobilayer are
disscussed. Finally, we conclude in Sec. \ref{Conc}.

\begin{figure}
\includegraphics[width=8.5cm]{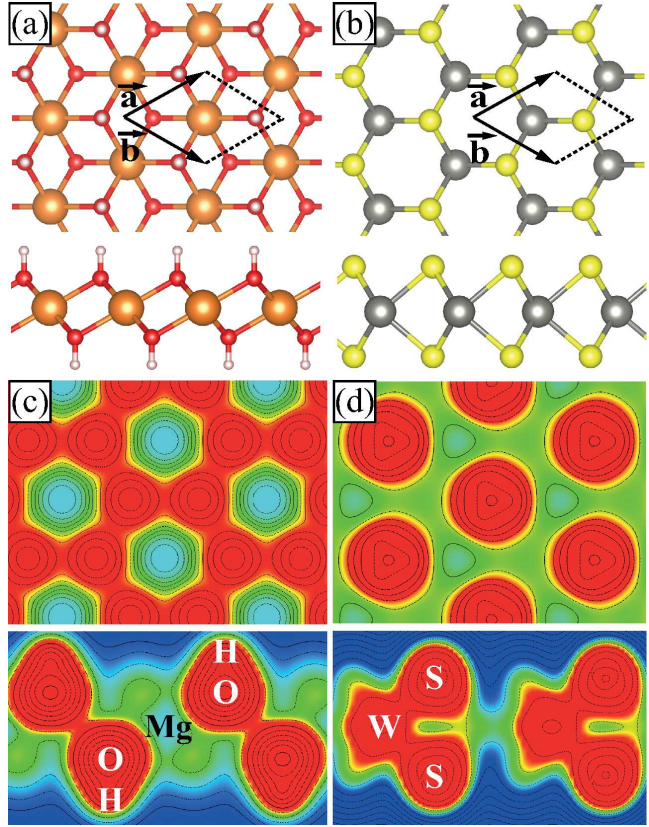}
\caption{\label{stuc}
(Color online) Top and side view of monolayers of (a) Mg(OH)$_2$ and (b)
WS$_2$. The charge distribution on the
individual atoms are shown in top and side views of (c) Mg(OH)$_2$ and (d)
WS$_2$.
Increasing charge density is shown by a color scheme from blue
to red with the formula F(N)=$1\times1000^{N/step}$ where step size taken to be
10 and N ranges from -1 to 2.}
\end{figure}

\section{Computational Methodology}\label{comp}

For our first-principles
calculations, we employed the
plane-wave basis projector augmented wave (PAW)
method in the framework of density-functional theory (DFT).
For the exchange-correlation potential, the generalized gradient approximation
(GGA) in the Perdew-
Burke-Ernzerhof (PBE) form\cite{GGA-PBE1,GGA-PBE2} was employed as implemented
in the Vienna
\textit{ab-initio}
Simulation Package (VASP)\cite{vasp1,vasp2}. The van
der
Waals (vdW) correction to the GGA functional was included
by using the DFT-D2 method of Grimme\cite{Grimme}. The inherent underestimation
of the band gap given by
DFT within the inclusion of spin-orbit-coupling (SOC) is corrected by using
the Heyd-Scuseria-Ernzerhof (HSE) screened-nonlocal-exchange functional of the
generalized Kohn-Sham
scheme\cite{Heyd}. Analysis of the charge transfers in the
structures was determined by the Bader
technique\cite{Henkelman}.

The dielectric function and the optical oscillator strength of
the individual monolayers and the heterostructure
were calculated by solving the Bethe-Salpeter equation (BSE) on top of single
shot GW (G$_{0}$W$_{0}$) calculation which was performed over standard DFT
calculations including
spin-orbit coupling (SOC). During this process we used $6\times6\times1$
$\Gamma$-centered \textbf{k}-point sampling. The cutoff for the response
function was set to 200 eV. The number of bands used in our calculations is
320.
The cutoff energy for the plane-waves was chosen to be 400 eV. We included
4
valence and 4 conduction bands into the calculations in the BSE step.

The
energy cut-off value for the plane wave basis set was taken to be
$500$ eV. The total energy
was minimized until the energy variation in successive steps became less
than $10^{-5}$ eV in the structural
relaxation and the convergence criterion for the
Hellmann-Feynman forces was taken to be $10^{-4}$ eV/\AA {}. The
minimum energy was
obtained by varying the lattice constant and the pressure was reduced below 1
kbar. 27$\times$27$\times$1 $\Gamma$-centered \textbf{k}-point sampling is used
for
the primitive unit cell.
The Gaussian broadening for the density of states calculation was taken to be
0.05. In order to investigate the
effect of an external electric field, an electric field is applied in
the direction normal to the plane of the heterobilayer. The binding energy per
unit
cell was calculated by
using the following formula:
$E_\textrm{bind}$=$E_\textrm{WS$_2$}$+$E_\textrm{Mg(OH)$_2$}$-$E_\textrm{hetero}
$,
where $E_\textrm{WS$_2$}$ and $E_\textrm{Mg(OH)$_2$}$ denote the total
energies of WS$_2$
and Mg(OH)$_2$ monolayers, respectively, while $E_\textrm{hetero
}$ denotes the
total
energy of the heterobilayer structure.

\begin{table*}
\caption{\label{table} The calculated ground state properties of
monolayer and their heterobilayer structures: structure,
 lattice
parameters of primitive unit cell, $a$ and $b$ (see Fig. \ref{stuc}), the
distance between the individual atoms contained in each monolayer $d_{X-Y}$,
magnetic state, the total
amount of charge recieved by the O or S atoms $\Delta\rho$, the binding
energy per unit cell between the monolayer in the heterobilayer $E_{bind}$, the
energy band gap of the
structure calculated within GGA ($E_\textrm{g}^{GGA}$), SOC
($E_\textrm{g}^{SOC}$) and HSE06
($E_\textrm{g}^{HSE}$), and workfunction $\Phi$ determined from Mg(OH)$_2$
side.}
\begin{tabular}{rcccccccccccccccc}
\hline\hline
& Geometry & $a$ & $b$ & $d_\textrm{Mg-O}$ & $d_\textrm{O-H}$ &
$d_\textrm{W-S}$ & Magnetic &
$\Delta\rho$
& $E_{\textrm{bind}}$ & $E_\textrm{g}^{GGA}$ & $E_\textrm{g}^{SOC}$ &
$E_\textrm{g}^{HSE}$ & $\Phi$ &\\
&   & (\AA{}{})  & (\AA{}{}) & (\AA{}{}) & (\AA{}{}) & (\AA{}{}) & State &
($e$) & (meV) & (eV) &
(eV) & (eV) & (eV) &\\
\hline
Mg(OH)$_2$ & 1T & 3.13 & 3.13 & 2.09 & 0.96 & - & NM & 2.9 & - & 3.25
&
3.22 & 4.75
& 4.15 \\
WS$_2$ & 1H & 3.18 & 3.18 & - & - & 2.41 & NM & 1.1 & - & 1.86 & 1.54 & 2.30 &
5.29
\\
Heterobilayer & 1T & 3.16 & 3.16 & 2.10 & 0.96 & 2.41 & NM & - &
147 &  1.05 &
0.97 & 2.24 &
4.34
\\
\hline\hline
\end{tabular}
\end{table*}

\section{single layer M\lowercase{g}(OH)$_2$ and WS$_2$}\label{monolayer}

Monolayer Mg(OH)$_2$ consists of hydroxyl (OH) groups bonded to Mg
atoms. As seen in Fig. \ref{stuc}, the
layer of Mg atoms is sandwiched between the OH groups in which O
and H atoms are strongly bonded to each other. The calculated lattice
parameters for monolayer Mg(OH)$_2$ are $a$=$b$=3.13 \AA {}. The thickness of
monolayer Mg(OH)$_2$ is 4.01 \AA {}.
The bond length of
Mg-O and O-H bonds are
calculated to be 2.09 \AA {} and 0.96 \AA {}, respectively. Bader charge
analysis shows that ionic bond character is present in the Mg(OH)$_2$
monolayer.
In
the structure each H atom donates 0.6 $e$ to neighboring O atom and each Mg
donates 0.85
$e$ per O atom.

Generic forms of monolayer structures of TMDs display honeycomb lattice
symmetry with the
1H phase for the dichalcogenides of Mo and W atoms. The  calculated lattice
parameters for the 1H phase of  WS$_2$ monolayer are $a$=$b$=3.18 \AA {} which
is very close
to that of Mg(OH)$_2$ monolayer. The W-S bond length in WS$_2$ is calculated
to be 2.42 \AA {}. The thickness of the layer is 3.13 \AA {} which is thinner
than monolayer Mg(OH)$_2$. In the monolayer WS$_2$ 0.55 $e$ of charge
accumulation occurs from a W atom to
each of the S atoms and the corresponding bonding character is covalent.

\begin{figure}[htbp]
\includegraphics[width=8.5cm]{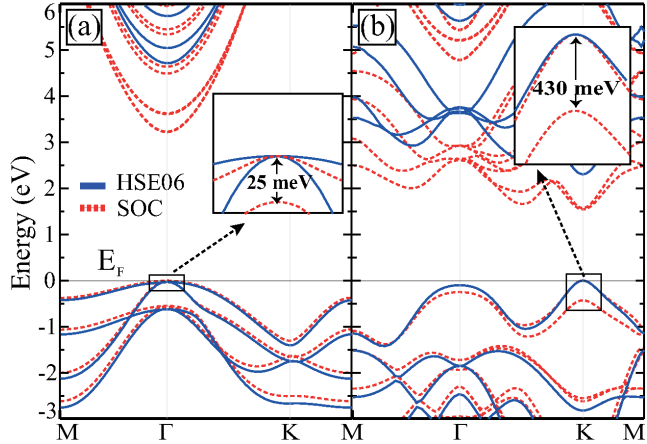}
\caption{\label{monoband}
(Color online) Calculated energy-band structure of monolayer (a) Mg(OH)$_2$ and
(b)
WS$_2$. The Fermi energy ($E_F$) level  is
set to the valence band maximum.}
\end{figure}

The calculated band structures within HSE06 correction are shown in
Fig. \ref{monoband}. Monolayer Mg(OH)$_2$ is found to be a direct band gap
semiconductor with a band gap of 4.75 eV. Both the valence band maximum (VBM)
and the conduction band minimum (CBM) reside at
the $\Gamma$ point in the Brillouin zone (BZ). The states in the VBM of the
Mg(OH)$_2$
monolayer are composed of $p_x$ and $p_y$ orbitals of the O atoms.

Similar to the monolayer Mg(OH)$_2$, monolayer WS$_2$ is also a direct band
gap semiconductor but with a lower band gap of 2.30 eV. As in other TMDs,
both the VBM and CBM of single layer WS$_2$ lie at the K point in the BZ.
As seen in Fig. \ref{monoband}(b), spin-orbit interaction at the VBM states is
much stronger since the states are
composed of $d_{x^2}$ and $d_{z^2}$ orbitals of W atoms. There is an
energy splitting of 430 meV at VBM which is much larger than that of monolayer
Mg(OH)$_2$ which is calculated to be 25 meV.

\section{Heterobilayer}\label{hetero}
\begin{figure}[htbp]
\includegraphics[width=8.5cm]{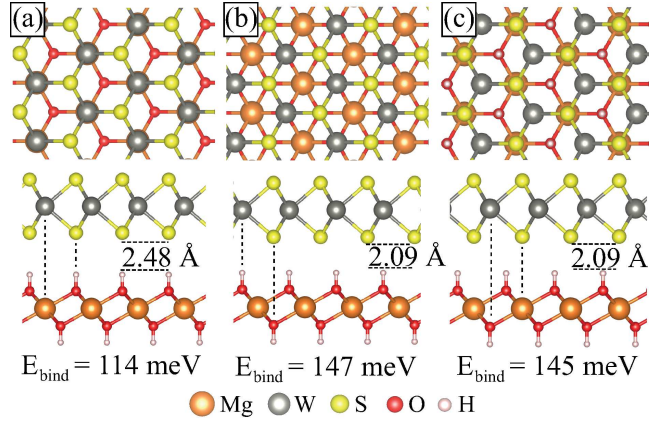}
\caption{\label{bilayer}
(Color online) Different possible stacking configurations for the heterobilayer
structure. (a) W atom on top of Mg atom, (b) W atom on top of upper OH group,
and (c) W atom on top of lower
OH group.}
\end{figure}

The calculated lattice
constants of Mg(OH)$_2$ and WS$_2$ monolayers are very close to each other and
therefore
it
is possible to construct a heterostructure of these monolayers where we may
assume a primitive unit cell containing 8 atoms in total. We considered three
different high-symmetry stacking configurations of the
monolayers (see Fig. \ref{bilayer}). We found that two of the stacking
configurations have
binding energies very close to each other but the one
with the W atoms residing on top of an interface OH group is the ground state
with a
binding energy of 147 meV. For the lowest energy stacking configuration the
interlayer distance is calculated to be 2.09
\AA {} and the individual atomic bond lengths remain the same as in their
isolated layers. The analysis for the charge transfers
between the
individual layers demonstrate that there is no depletion from one layer to the
other for all the stacking geometries shown in
Fig.
\ref{bilayer}. This result is expected due to the weak vdW interaction
between the individual layers.

\begin{figure}
\includegraphics[width=8.5cm]{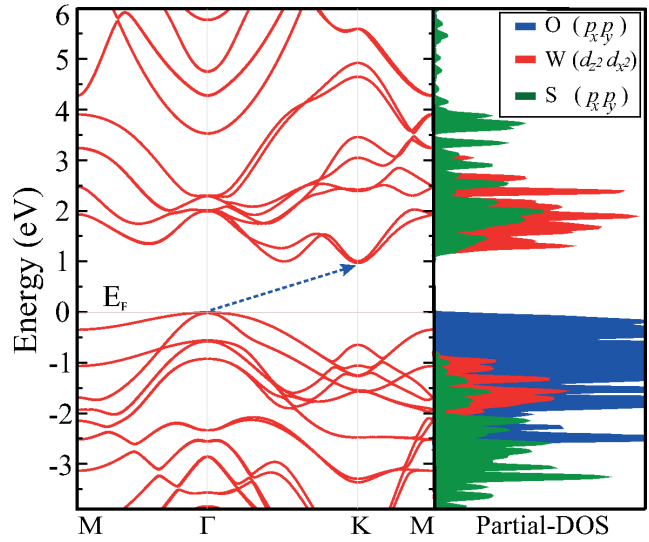}
\caption{\label{bands}
(Color online) The
band structure (left) and the corresponding partial density of states (PDOS)
(right) of the
heterobilayer structure calculated within SOC. The Fermi energy ($E_F$) level
is
set to the valence
band maximum.}
\end{figure}

The calculated energy-band structure for the heterobilayer displays a
semiconducting character with an indirect band gap of
2.24 eV. As seen in Fig. \ref{bands}, the VBM of the heterobilayer that
originates from the Mg(OH)$_2$ layer lies at the $\Gamma$ point while the CBM of
the structure which arises from the WS$_2$ layer lies at the K point. Calculated
energy-band diagram of the heterostructure also indicates the weak interlayer
interaction. As seen in Fig. \ref{bands}, the partial DOS
(PDOS) indicates that the VBM of heterobilayer exclusively consists of $p_x$
and $p_y$
orbitals of the O atoms while the CBM is characterized by the orbitals of W and
S atoms. This also demonstrates the type-II nature of the heterojunction: the
two band edges originate from different
individual layers and consequently the excited electrons and holes are confined
in
different
layers which leads to the formation of spatially indirect excitons.

\begin{figure}[htbp]
\includegraphics[width=8.5cm]{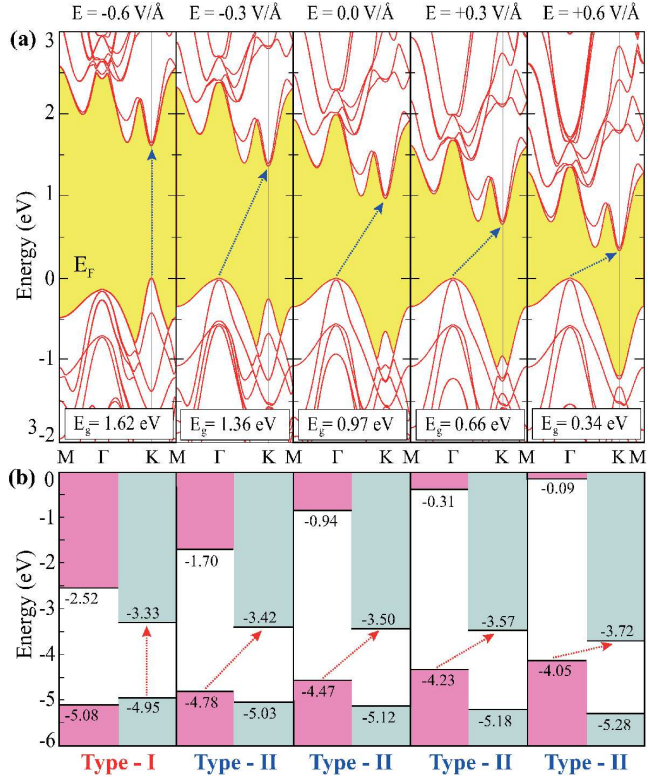}
\caption{\label{field}
(Color online) (a) The effect of an external out-of-plane electric field on the
band
structure of the heterobilayer and (b) the corresponding band alignments (vacuum
levels are set to zero). The
band gap regions are highlighted
in
yellow while the CBM and VBM are highlighted in pink and grey for Mg(OH)$_2$
and WS$_2$, respectively.}
\end{figure}

\section{Effect of External Electric Field}\label{Efield}
Applying an external electric field is one of the common method to modify or
tune the
physical properties of
materials. In the field of 2D materials, a perpendicular electric field can
lead to doping and in the case of bilayers
it can induce charge transfer between layers. Castro \textit{et
al.} reported that the electronic band gap of a graphene bilayer structure can
be controlled externally by applying a gate bias. They showed that
the band gap changes from zero to midinfrared energies for field values
$\leq$ 1 V/nm\cite{Castro}. Chu \textit{et
al.} showed a continuous
bandgap tuning in bilayer MoS$_2$ with applied gate voltage\cite{Chu}. Here we
present our results for the effect of a perpendicular electric field on
the electronic and optical properties of the heterobilayer.

As seen in Fig. \ref{field}(a), the heterostructure is an indirect band gap
semiconductor when there is no external
electric field, in which the VBM is at $\Gamma$ but the CBM is at the K point.
Appyling a positive electric field decreases the band gap (from 0.97 eV to 0.34
eV
for E= +0.6 V/\AA {}). The reason for such decreasing band gap is the
shift of the band edges at the $\Gamma$ and the K points. Increasing the value
of the
positive electric field shifts the VBM of Mg(OH)$_2$ up in energy while it
shifts the CBM of WS$_2$ down resulting in a decrease of the energy gap. The
indirect
character
of the energy gap is not affected by the field.
However, changing the direction of the applied electric field widens the band
gap and ultimately leads to an indirect-to-direct band-gap-crossover as seen in
Fig.
\ref{field}(a).

When the strength of the electric field is -0.6 V/\AA {}, it is
clearly seen that both VBM and CBM of the heterobilayer reside at the K high
symmetry point in the BZ. Thus, a transition from staggered gap to a straddling
gap (type-I
heterojunction) occured as shown in Fig. \ref{field} (b).
In fact, the critical electric field value for which this indirect-to-direct
band-gap-crossover occurs is calculated to be 0.51 V/\AA {}.
At this critical value of the applied electric field, the valence band edge
energy of the bands at the $\Gamma$ and K points become degenerate.
As seen in Fig. \ref{field}(a), the bands at the valence band edge of the K
point, which originate from the WS$_2$ layer, shift up while the bands at the
$\Gamma$ point which originate from the Mg(OH)$_2$ layer shift down when making
the external electric field more negative. Due to these opposite
shifts of the VBM of the individual layers (see Fig. \ref{field}(b)) a
transition from
indirect-to-direct gap is predicted at a certain value of the applied
field. After the transition to type-I
heterojunction both type of charge carriers are confined to the WS$_2$ layer
which is desirable for
applications in optoelectronic
devices and for semiconductor laser applications. It is also
important to point out that including quasiparticle energies might slightly
change
the
band gap and the electric field value for which the indirect-to-direct
band-gap-crossover occurs. However, the overall tunability characteristic of
the
heterobilayer using electric field would remain the same.

\section{OPTICAL PROPERTIES}\label{optical}

\begin{figure}[htbp]
\includegraphics[width=6.5cm]{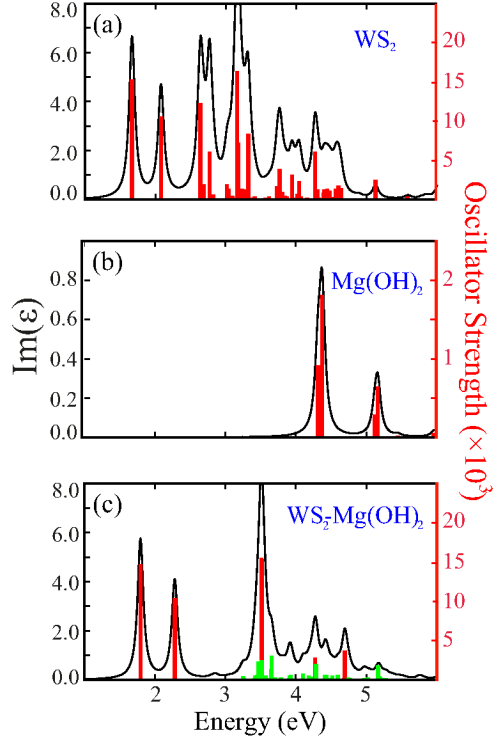}
\caption{\label{diel}
(Color online) Imaginary part of the dielectric function and the
oscillator strength of the optical transitions of (a) WS$_2$ monolayer (b)
Mg(OH)$_2$ monolayer and (c) WS$_2$-Mg(OH)$_2$ heterostructure. The oscillator
strength of the optical transitions shown by green and red correspond
to inter and intralayer optical transitions, respectively.}
\end{figure}

\begin{figure*}[htbp]
\includegraphics[width=16.0cm]{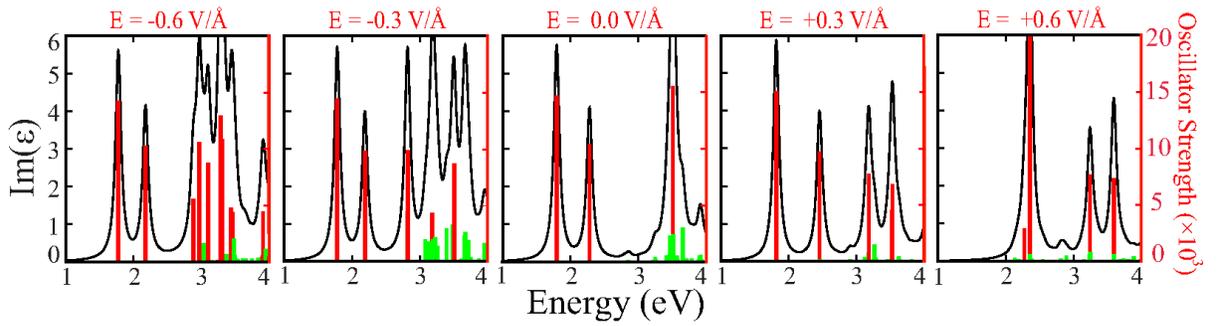}
\caption{\label{diel2}
(Color online) Imaginary part of the dielectric function and the
oscillator strength of the optical transitions for different values of the
perpendicular electric field. The oscillator strength of the
optical transitions which are shown in green and red correspond to
inter- and intralayer optical transitions in the heterostructure, respectively.}
\end{figure*}

In order to investigate the optical properties of the isolated monolayers and
the
heterostructure, we solved the
BSE equation on top of G$_0$W$_0$
calculation.
In Fig.~\ref{diel} we show the imaginary part of the dielectric
function and the oscillator strength of the optical transitions of WS$_2$
(Fig.~\ref{diel}(a)), Mg(OH)$_2$ (Fig.~\ref{diel}(b)) and
WS$_2$-Mg(OH)$_2$ heterostructure (Fig.~\ref{diel}(c)).

The first two peaks at 1.67 and 2.08 eV in the optical spectrum of monolayer
WS$_2$ (Fig.~\ref{diel}(a)) originate from the  optical transitions
at the K point in the BZ. The splitting (410 meV) of these two peaks
is consistent with the splitting of the VBM bands at the K point due to the SOC
effect (Fig.~\ref{monoband}).
Our calculations show that the oscillator strength of the optical transitions of
the WS$_2$ monolayer is
an
order of magnitude larger than that of the Mg(OH)$_2$ monolayer. The first two
optical transitions for the monolayer Mg(OH)$_2$ (Fig.~\ref{diel}(b)) are split
with a very small
energy of 58 meV, this value is close to the value of the VBM splitting at the
$\Gamma$ point due to the SOC effect (Fig.~\ref{monoband}).

In Fig.~\ref{diel}(c) we show the imaginary part of
the dielectric function and the oscillator strength of the optical transitions
of the WS$_2$-Mg(OH)$_2$ heterostructure. As seen from the figure the first
two peaks in the optical spectrum originate from the WS$_2$ monolayer. The
splitting of the first two peaks increases to 480 meV and the positions of them
are blueshifted
due to the interaction between the two monolayers.
Although the oscillator strength of the peaks from the Mg(OH)$_2$ monolayer are
small, they can still be identified around 4 eV in the optical spectrum.
The exciton binding energy of WS$_2$,  Mg(OH)$_2$ and the heterobilayer is
found to be as 0.84 eV, 2.4 eV and 0.74 eV, respectively.

As discussed earlier, the WS$_2$-Mg(OH)$_2$ heterostructure is
a type-II heterojunction in the absence of an electric field. In this kind of
heterojunctions spatially direct absorption (intralayer
excitons) and spatially indirect emission (intralayer excitons) are expected as
observed experimentally in the WSe$_2$-MoS$_2$ heterostructure. \cite{fangpnas}
Therefore, different exciton peaks might dominate the optical spectrum
depending on the measurement method.
For instance, the intralayer excitons (red optical transitions in
Fig.~\ref{diel}(c)) will dominate the absorption spectrum, while interlayer
excitons (green optical transitions in
Fig.~\ref{diel}(c))
dominate the emission spectrum (i.e. photoluminescence (PL) measurements) of
the
heterostructure. So, in order to make a reasonable comparison between
experiment and our
calculations, we identified the optical transitions that correspond to inter-
and
intralayer excitons and plot them, respectively, in green and red color in Fig.
\ref{diel}(c). As shown in Fig.~\ref{diel}(c), the main
optical transitions of the heterostructure originate from the intralayer
recombinations and their oscillator strength
is an order of magnitude larger than the oscillator strength of the interlayer
excitons.

As mentioned before, the WS$_2$-Mg(OH)$_2$ heterobilayer has a type-II
alignment and it transforms into a type-I heterostructure under an external
out-of-plane electric field of -0.51 V/\AA {}.
In order to investigate the variation in the optical properties of the
heterobilayer under different electric field strengths, we calculated the
dielectric
function and the
oscillator strength of the different optical transitions under out-of-plane
electric
field of -0.6, -0.3, 0 , 0.3 and 0.6 V/\AA {} which are shown in
Fig.~\ref{diel2}. Since the
WS$_2$-Mg(OH)$_2$ heterobilayer
remains type-II for out-of-plane electric field  of -0.3, 0 ,
0.3 and 0.6 V/\AA, the red optical transitions will dominate the absorption
spectrum but
the green ones dominate the PL measurements. When the
external out-of-plane electric field is -0.6 V/\AA {},
the structure becomes type-I. In this case, the intralayer
optical transitions (red) will dominate the PL measurements.
Therefore, we predict an increase in the PL intensity of the WS$_2$-Mg(OH)$_2$
heterobilayer when the out-of-plane electric field becomes more negative than
-0.51 V/\AA {}. Applying perpendicular electric field
also modifies the exciton
binding energy of the heterobilayer. According to our calculations, the exciton
binding energy of the heterobilayer becomes 0.77, 0.76, 0.65 and 0.10 eV for
-0.6, -0.3, 0.3 and 0.6 V/\AA {} electric field, respectively.

\section{Conclusion}\label{Conc}

We investigated the structural, electronic and optical properties of the
monolayers
Mg(OH)$_2$ and WS$_2$ and its
heterobilayer structure. In addition the effect of an applied out-of-plane
electric field on the electronic and optical properties of the
heterobilayer were investigated.
We found
that both Mg(OH)$_2$ and WS$_2$ are direct-gap
semiconductors while the Mg(OH)$_2$-WS$_2$
heterobilayer structure
is an indirect-gap semiconductor. Our results
demonstrated that both
the band gap and the energy-band dispersion of the heterobilayer
structure can be tuned by the application of an external perpendicular electric
field. At an applied electric field of -0.51 V/\AA {} a transition from a
staggered to a straddling gap heterojunction occurs which can be used
for optoelectronic
and semiconductor laser applications. In addition, by solving the
Bethe-Salpeter equation on top of single shot
G$_0$W$_0$ calculations, we predict that the oscillator strength of the
intralayer excitons of the heterostructure is an order of
magnitude larger than the oscillator strength of the interlayer excitons. It
appears that heterobilayers of TMDs and AEHs may find applications in various nanoscale optoelectronic devices.

\begin{acknowledgments}

This work was supported by the Flemish Science Foundation (FWO-Vl) and the
Methusalem foundation of the Flemish government. Computational resources were
provided by TUBITAK ULAKBIM, High Performance and Grid Computing Center
(TR-Grid e-Infrastructure). H.S. is supported by a FWO Pegasus Long Marie Curie
Fellowship. H.S. and R.T.S. acknowledge the support from
TUBITAK through project 114F397.

\end{acknowledgments}

\end{document}